\documentclass[twocolumn,twoside]{IEEEtran}

\usepackage{graphics} 
\usepackage{epsfig} 
\usepackage{epstopdf}
\usepackage{color,soul}
\usepackage{amsmath,graphicx}
\usepackage{multirow}
\usepackage{algorithmic}
\usepackage{algorithm}
\usepackage{float}
\usepackage[caption = false]{subfig}
\usepackage{paralist}
\usepackage{amsmath,amssymb,amstext} 
\usepackage{color}
\usepackage{lipsum}

\begin{document}
\title{\LARGE \bf Noise-Compensated, Bias-Corrected Diffusion Weighted Endorectal Magnetic Resonance Imaging via a Stochastically Fully-Connected Joint Conditional Random Field Model}

\author{Ameneh Boroomand$^*$,~\IEEEmembership{Student Member,~IEEE}, Mohammad Javad Shafiee,~\IEEEmembership{Student Member,~IEEE}, Farzad Khalvati,~\IEEEmembership{Member,~IEEE}, Masoom A. Haider, Alexander Wong,~\IEEEmembership{Senior Member,~IEEE}
\thanks{A.~Boroomand$^*$, M.~J.~Shafiee and A.~Wong are with the Department of Systems Design Engineering, University of Waterloo, Waterloo, Ontario, Canada, N2L 3G1. {\tt\small aborooma@uwaterloo.ca}}
\thanks{F.~Khalvati and M.~A.~Haider are with the Department of Medical Imaging, University of Toronto, and Sunnybrook Research Institute, Toronto, Ontario, Canada, M4N 3M5. {\tt\small farzad.khalvati@sri.utoronto.ca}}

\thanks{Copyright (c) 2016 IEEE. Personal use of this material is permitted. However, permission to use this material for any other purposes must be obtained from the IEEE by sending a request to pubs-permissions@ieee.org.}
}
\maketitle
\markboth{IEEE Transactions on Medical Imaging}
{Boroomand \MakeLowercase{\textit{et al.}}:Noise-Compensated, Bias-Corrected Endorectal Diffusion Imaging.}

\IEEEpeerreviewmaketitle

\begin{abstract}
Diffusion weighted magnetic resonance imaging (DW-MR) is a powerful tool in imaging-based prostate cancer screening and detection. Endorectal coils are commonly used in DW-MR imaging to improve the signal-to-noise ratio (SNR) of the acquisition, at the expense of significant intensity inhomogeneities (bias field) that worsens as we move away from the endorectal coil. The presence of bias field can have a significant negative impact on the accuracy of different image analysis tasks, as well as prostate tumor localization, thus leading to increased inter- and intra-observer variability. Retrospective bias correction approaches are introduced as a more efficient way of bias correction compared to the prospective methods such that they correct for both of the scanner and anatomy-related bias fields in MR imaging. Previously proposed retrospective bias field correction methods suffer from undesired noise amplification that can reduce the quality of bias-corrected DW-MR image. Here, we propose a unified data reconstruction approach that enables joint compensation of bias field as well as data noise in DW-MR imaging. The proposed noise-compensated, bias-corrected (NCBC) data reconstruction method takes advantage of a novel stochastically fully connected joint conditional random field (SFC-JCRF) model to mitigate the effects of data noise and bias field in the reconstructed MR data. The proposed NCBC reconstruction method was tested on synthetic DW-MR data, physical DW-phantom as well as real DW-MR data all acquired using endorectal MR coil. Both qualitative and quantitative analysis illustrated that the proposed NCBC method can achieve improved image quality when compared to other tested bias correction methods. As such, the proposed NCBC method may have potential as a useful retrospective approach for improving the consistency of image interpretations.
\end{abstract}


\section{Introduction}
Prostate Cancer (PCa) has been shown to be the most prevalent type of cancer among the male population all around the world. Based on statistics from the Canadian Cancer Society, an estimated 24,000 Canadian were diagnosed with PCa in 2015, while 4,100 men were expected to die of PCa in 2015~\cite{Cancerstats}. Due to the high rate of PCa incidences, there is an urgent need for developing more reliable and accurate PCa diagnostic and screening tools as well as optimizing the current conventional PCa detection methods to help with the early diagnosis and treatment of the disease.

Several primary clinical tests are usually recommended as routine checks for individuals who are suspected of having PCa. The most routine clinical examination is a simple blood test which determines the level of prostate-specific antigen (PSA)~\cite{partin1993use} in a patient's blood, as an elevated PSA level is often observed in men with PCa. However, the results of PSA tests are often unreliable due to dependency of the PSA level on several uncontrolled factors~\cite{catalona1994comparison}. Digital Rectal Examination~\cite{smith1995interexaminer} is a routine physical test used for the initial examination of the prostate gland. The sensitivity of this method is highly dependent on the expertise of the physician who performs the examination and, therefore, it can often be unreliable for the early diagnosis of PCa~\cite{cooner1990prostate}. Trans-rectal ultrasound (TRUS) biopsy is another routine test that is shown to have high specificity for the PCa diagnosis. However, the low sensitivity of this technique may result in misdiagnosis and failure of some high grade cancers and consequently undertreatment of PCa candidates~\cite{blomqvist2014limited}. Magnetic resonance (MR) imaging has been shown as one of the most reliable imaging techniques for early diagnosis and screening of PCa candidates~\cite{lawrentschuk2013role}. Diffusion weighted MR (DW-MR) imaging is established as a promising MR modality that can be useful for the aim of PCa detection along with the other MR modalities such as T2-weighted MR images~\cite{hosseinzadeh2004endorectal,haider2007combined,tamada2014diffusion}. In addition, multi-parametric MRI (mpMRI) is becoming an important screening tool for prostate cancer screening through which unique information is captured using different imaging modalities such as DW-MR and T2w enabling a more accurate early detection of PCa~\cite{arumainayagam2013multiparametric}.

The DW-MR acquisition process is typically parameterized by the $b$-value, which characterizes the amplitude, duration and temporal spacing of the DW-MR gradient pulses~\cite{bammer2003basic}. Previous studies show that DW-MR acquisitions with higher $b$-values result in greater contrast between the cancerous and benign tissues. However, the main challenge in using DW-MR imaging with higher $b$-values is a significant reduction in signal-to-noise ratio (SNR)~\cite{kim2010high,katahira2011ultra,ueno2013ultra}.

To improve SNR for DW-MR acquisitions to obtain adequate signal, a commonly-used solution is to make use of a specific endorectal MR coil that is placed in the patient's rectum~\cite{hosseinzadeh2004endorectal,nakashima2004endorectal}. Although the endorectal coil is no longer considered a routine procedure for every prostate MRI exam and for many indications especially at 3T its use has been discontinued, it continues to be an important part of the mpMRI landscape. In many clinical centers, it continues to be used and is advocated in the following scenarios:
\begin{enumerate}
	\item In older 1.5T MRI systems where SNR is inadequate (still the largest installed base), the boost in SNR achieved using endorectal coil has led its use by many when looking for extraprostatic extension of PCa near the neurovascular bundles and seminal vesicles, both of which are directly near the receiver antenna when the endorectal coil is inserted.
\item Furthermore, endorectal coil becomes a necessity for prostate DW-MR acquisitions for obese patients with large abdominal girth, where the external coils are further away from the prostate and as a result SNR suffers significantly.
	\item In many centers, endorectal 3T MRI is still considered the best test for detecting small tumors with confidence or local staging of high importance~\cite{Heijmink}.
\end{enumerate}	

While endorectal coil can significantly boost SNR in DW-MR acquisitions, one of the drawbacks is that it also introduces significant intensity inhomogeneities, commonly referred to as bias field, in the prostate gland region, especially at the zones closer to the coil~\cite{viswanath2012central}. Such intensity inhomogeneities can negatively affect the performance of image analysis tasks such as segmentation and registration, and pose challenges for accurate interpretation and analysis of PCa, thus leading to increased inter- and intra-observer variability~\cite{viswanath2011empirical,liney1998simple}. Hence, there exists a need for bias correction methods to compensate for the significant intensity inhomogeneities caused by endorectal coil.

In addition to endorectal coil causing intensity inhomogeneities, non-endorectal DW-MR is also subject to artifacts such as intensity distortions, due to field inhomogeneity and differences in magnetic susceptibility in the imaged regions~\cite{chilla2015diffusion}, which can be potentially corrected for by the proposed NCBC algorithm.

Compensating for intensity inhomogeneities caused by endorectal coil can also benefit computer-aided diagnosis (CAD) tools. For example, several radiomics-based CAD algorithms have been proposed for automatic prostate cancer detection which use multi parametric MRI to extract texture and morphological features fed into a classifier~\cite{FKhalvati2015,KhalvatiMAPS}. Intensity inhomogeneities may cause two similar regions of interest have different textural characteristics which may lead to different features and hence, inaccurate classification of healthy and cancerous regions~\cite{Roy}.

In this paper, a unified noise compensated, bias-corrected (NCBC) data reconstruction method is proposed for enabling noise-free, bias-free endorectal DW-MR imaging of the prostate gland. To the best of the authors' knowledge, previous studies have not provided solutions for simultaneous bias correction and data noise compensation within a unified framework. The proposed NCBC framework is designed in a Maximum a Posteriori (MAP) framework which takes advantage of a novel stochastically fully-connected joint conditional random field (SFC-JCRF) model to find an optimal estimate of the noise-free, bias-free endorectal MR data. The utilized SFC-JCRF model incorporates different contextual information for improving the quality of the reconstructed DW-MR data and better adoption to different tissue types when it takes advantage of stochastic cliques to model the long-range spatial interactions among the image pixels.

\section{Related Works}
To deal with the presence of bias field, most current MR scanners are integrated with built-in Surface Coil Intensity Correction (SCIC)~\cite{avinash1999method} designed to mitigate intensity inhomogeneities in the data acquisitions. Different prospective and retrospective methods have been applied to reduce the effect of bias fields in different MR imaging modalities as well as for various organs~\cite{vovk2007review,likar2001retrospective}. Prospective methods only compensate for the scanner-related bias fields and do not account for the possible bias fields that result due to the patient's anatomy. Prospective approaches mainly take advantage of a MR phantom~\cite{collewet2002correction}, multi-coil MR imaging~\cite{brey1988correction,narayana1988compensation}, or specific designed MR imaging sequences~\cite{mihara1998method} for the purpose of bias correction.

Retrospective methods are more efficient and flexible as they can compensate for both scanner- and anatomy-related bias field artifacts by taking advantage of information from the acquired data~\cite{vovk2007review}. Filter-based methods~\cite{brinkmann1998optimized}, surface fitting methods~\cite{dawant1993correction,tincher1993polynomial,brechbuhler1996compensation,styner2000parametric}, segmentation-based methods~\cite{ahmed2002modified,xie2015modified,li2011level,li2014multiplicative} as well as histogram-based methods~\cite{mangin2000entropy,milles2007mri} are the most well-known retrospective approaches for bias correction. The advantages and disadvantages of different bias correction methods are fully discussed in the related literature~\cite{velthuizen1998review,vovk2007review}. In general, filter-based methods assume that the MR bias field are largely characterized at the low frequencies and, therefore, can be suppressed using a low-pass filter. Filter-based methods in general are fast and simple to implement and deploy, but may result in the loss in useful low frequency content~\cite{velthuizen1998review}. Bias correction using surface fitting methods rely on the modeling of bias field as a parametric function such as a polynomial or spline, and the performance of this approach is highly dependent on the selection of true initial points as well as initial modeling of bias field~\cite{dawant1993correction,styner2000parametric,vovk2007review}. Segmentation-based methods perform alternating bias correction and segmentation steps, with the accuracy of each step having a large effect on the other. Furthermore, this approach can be time consuming, especially if the goal is only for bias correction. Histogram-based methods mainly take advantage of the image intensity histogram or image entropy to perform the bias correction task~\cite{vovk2007review}. These methods usually assume a prior distribution for modeling each of the specific tissue such that it may limit effectiveness when used with other tissue types~\cite{hui2010fast,jagercorrection}.

A common challenge faced in bias correction, particularly when dealing with the significant intensity inhomogeneities exhibited in DW-MR acquisitions using endorectal coil, is the undesired noise amplification which can have significant negative effects on not only interpretation but also tasks such as segmentation and registration~\cite{gelber1994surface,liney1998simple}. Most existing bias correction methods either do not take into account the issue of noise amplification~\cite{xie2015modified,li2014multiplicative,han2001multi,lui2014monte} or add a simple spatial filter to the designed method to control the effect of data noise as an independent pre- or post-processing step~\cite{salvado2006method,wang2014bias}. To better handle noise amplification due to bias correction, in one of our recent works~\cite{lui2015monte} a Monte Carlo-based denoising algorithm was proposed as a post-processing step that aims to suppress noise on bias-corrected endorectal MR images.  However, such post-processing methods that focus purely on denoising are limited in their ability to deal with the noise amplification due to the fact that the imaging data has been modified independently using a separate bias-correction method, and as such the inherent noise characteristics have been altered in a nonstationary manner that is difficult to model accurately.  Therefore, a different strategy that can jointly account for and compensate bias field effects and MR noise jointly in the same model, such as the proposed NCBC framework is highly desired.

The concept of random fields has been used in a few segmentation-based bias correction methods which simultaneously performs the task of tissue segmentation and bias field correction~\cite{van1999automated,xie2015modified,ji2012brain}. Markov random field (MRF) models have also been utilized for the purpose of MR noise suppression~\cite{awate2007feature}. The nature of random field modeling allows for the incorporation of different contextual information in the designed segmentation-based bias correction or noise suppression algorithm such that it can improve the effectiveness of these methods to better adapting to different tissue types~\cite{awate2007feature}. 

\section{Methodology} \label{Methodology}

\label{sec:method}
The goal of the proposed noise-compensated, bias-corrected (NCBC) data reconstruction method is to estimate the noise-free, bias-free endorectal DW-MR data based on a novel conditional random field model that we call the stochastically fully connected joint conditional random field (SFC-JCRF) model. Conditional random field is a form of undirected graphical model first introduced for data labeling task~\cite{lafferty2001conditional}, while it has been incorporated in several applications  such as image segmentation, classification and object detection~\cite{panjwani1995markov,plath2009multi,duvenaud2011multiscale,quattoni2004conditional} and it has demonstrated promising results. Conditional random field models can handle different dependencies that exist among the observations and states as they directly model the conditional probability of the states given a set of observations. The direct modeling of the conditional probability of states given observations relaxes the conditional independence assumption that is required in Markov random field models~\cite{lafferty2001conditional}. Here, we formulate the proposed method within a maximum a posterior (MAP) framework that leverages a SFC-JCRF model to define a joint conditional probability for the MR bias field and data noise, which will be explained in detail below.
\vspace{-0.1in}
\subsection{Problem Formulation}
Let $S$ be a set of pixels in a discrete lattice $\mathcal{L}$ upon which a DW-MR image is defined, and $s \in S$ be a pixel in $\mathcal{L}$. A DW-MR acquisition corrupted by intensity inhomogeneities (bias field) and data noise can be modeled as the following forward problem:
\begin{equation}
V=M\cdot B+\mathcal{N},
\label{eq1}
\end{equation}
\noindent where \mbox{$V=\{v_s|s\in S\}$}, \mbox{$B=\{b_s|s\in S\}$} and \mbox{$M=\{m_s|s\in S\}$} are random fields characterizing the DW-MR acquisition, bias field, and the noise-free, bias-free endorectal DW-MR image, respectively where $s \in S$ encodes a node in the underlying graph $S$. In this work, $\mathcal{N}$ accounts for the inherent data noise present in the DW-MR acquisition which is modeled using Rician distribution. Based on the forward problem, the goal of the proposed NCBC method is to estimate the noise-free, bias-free endorectal DW-MR image $M$ by solving the inverse problem, which we can formulate within a MAP framework as
\begin{equation}
\hat{M}=\underset{M \in \mathcal{M}}{argmax} P(M,B|V).
\label{eq2}
\end{equation}
Here, $P(M,B|V)$ represents a joint posterior distribution defined over the bias field $B$ and noise-free, bias-free endorectal DW-MR image $M$, with $\mathcal{M}$ encoding the set of possible noise-free, bias-free endorectal DW-MR images. To solve this inverse problem, the proposed NCBC method takes advantage of a novel extension to the model proposed in~\cite{shafiee2014efficient}, which we term it as SFC-JCRF model.

In this model, both bias field $B$ and noise-free, bias-free endorectal DW-MR image $M$ are assumed as different individual random fields and modeled using a separate layer with connections to the observation layer $V$. Using CRF modeling~\cite{lafferty2001conditional}, the joint posterior distribution in Eq.~\ref{eq2} can be modeled as
\begin{equation}
P(M,B|V)=\frac{1}{Z(V)}\exp \Big(-E(B,V,M)\Big),
\label{eq3}
\end{equation}
\noindent where $Z(V)$ is the normalization constant and E$\left(\cdot\right)$ represents the energy function formulated as
\begin{align}
E(B,V,M)=\sum_{s \in S}\psi_{u}(b_{s,},m_{s},V)+
\sum_{\varphi\in C}\psi_{p}(b_{\varphi},m_{\varphi},V),
\label{eq4}
\end{align}
\noindent which is defined over the random fields of the bias field $B$, DW-MR acquisition $V$, and the noise-free, bias-free endorectal DW-MR image $M$. The energy function $E(\cdot)$ is the combination of unary $\psi_{u}\left(\cdot\right)$ and pairwise $\psi_{p}\left(\cdot\right)$ potential functions that enforce the likelihood and prior information into the model, with $C$ encoding a set of defined clique structures. Here, $b_\phi \in B$ and $m_\phi \in M$ are subsets of nodes in the clique $\varphi$ and $V$ is the total observation set (i.e. here the DW-MR acquisition). In the conventional CRF models, each node (pixel) has a local connection with its four or eight nearest neighbors while longer range interactions are ignored due to their computational complexity. However, taking into account longer range interactions could be useful for a better compensation of data noise and bias present in the DW-MR acquisition. To account for the longer range data interactions, the proposed SFC-JCRF model basically takes advantage of stochastic clique indicator functions with predefined probability distribution to determine the possible constitution of clique structure between all node pairs.

The unary and pairwise potential functions are defined as
\begin{align}
\psi_{u}(b_{s,},m_{s},V) &=\sum_{s=1}^{|M|}\alpha^{u} \cdot f(b_{s},m_{s},V)\\
\psi_{p}(b_{\varphi},m_{\varphi},V)&=\sum_{k = 1}^K \sum_{\{y_{s},y_{s'}\}\in y_{\varphi}}\alpha_{k}^{p} \cdot g_{k}(y_{s},y_{s'},h_{\varphi},V)
\end{align}
\noindent where $\alpha^{u}$ and $\alpha^{p}$ define the weights of unary and pairwise terms, $f(\cdot)$ and $g_{k}(\cdot)$ are the employed unary and pairwise feature functions, and $K$ represents the total number of pairwise feature functions. Here we consider the acquired DW-MR acquisition $V$ as the unary function and  a simple similarity function to enforce the smoothness constraint on the neighbor pixels is obtained as the pairwise function.

To find the best estimate of the noise-free, bias free endorectal DW-MR image $M$, the joint posterior distribution $P(M,B|V)$ should be maximized. The bias field $B$ can be jointly estimated through solving of the optimization problem in Eq.~\ref{eq2}.

\subsection{Inference}
To find the best estimate of noise-free, bias-free endorectal DW-MR image $M$, Eq.~\ref{eq2} is maximized by minimizing the energy function $E(B,V,M)$ in Eq. \ref{eq4}. To this aim, a gradient descent algorithm was applied to solve the optimization problem of Eq.~\ref{eq2}, where the noise-free, bias-free endorectal DW-MR image $M$ as well as bias field $B$ can be simultaneously estimated as,
\begin{equation}
M^{t+1}=M^{t}-\mu_{1}\frac{\nabla E(B,M,V)}{\nabla M}\; \; s.t.\;\; \text{$B$ is fixed}
\label{eq7}
\end{equation}
\begin{equation}
B^{t+1}=B^{t}-\mu_{2}\frac{\nabla E(B,M,V)}{\nabla B}\;\; s.t.\;\; \text{$M$ is fixed}
\label{eq8}
\end{equation}
where $\mu_1$ and $\mu_2$ are learning rates, $\frac{\nabla E(\cdot)}{\nabla B}$ and $\frac{\nabla E(\cdot)}{\nabla M}$ are the gradient of energy regarding the $B$ and $M$ respectively. It can be seen that this approach allows the bias field $B$ and noise-free, bias-free endorectal DW-MR image $M$ to be optimally estimated in an interchanging fashion. In formulations of Eq. \ref{eq7} and Eq. \ref{eq8}, $M^{t+1}$ and $B^{t+1}$ correspondingly denote the solutions of the optimization problem at $(t+1)^{th}$ iteration. To find the final solution, the value of noise-free, bias-free endorectal DW-MR image is estimated at a pixel level such that,
\begin{align}
\label{eq9}
&m_{s}^{t+1}=\\ \nonumber
&m_{s}^{t}-\rho\left(\frac{\partial\psi_{u}(b_{s,},m_{s},V)}{\partial m_{s}}\right)-\eta\left(\sum_{\varphi\epsilon C}\frac{\partial\psi_{p}(b_{\varphi},m_{\varphi},V)}{\partial m_{s}}\right),
\end{align}
\noindent where the bias field is assumed fixed. Here, $\theta = [\rho ,\eta]$ are the learning rates associated with unary and pairwise gradient terms. By fixing $M$ in each iteration after its optimization, the value of bias field $B$ at each pixel is calculated using the similar expression in Eq. \ref{eq9} and when the gradient was calculated with respect to $b_s$.

\section{Experimental Setup} \label{Experimental Setup}
\label{sec:exp}
All DW-MR imaging was performed with a 1.5 T MR  scanner at Sunnybrook Health Sciences Center, Toronto, ON, Canada and using a typical endorectal receiver coil (Medrad eCoil). All data was obtained retrospectively under the local institutional research ethics board. For the experiments, the DW-MR data acquisitions of seven different patient cases (four with prostate tumors and three without tumors) were obtained at three different b-values. All imaging experiments were performed according to a predefined standard imaging protocol with the imaging parameters shown in Table~\ref{Ta1}.


\begin{table}[ht]
	\renewcommand{\arraystretch}{1.3}
	\caption{endorectal imaging protocol}
	\label{Ta1}
	\centering
	\begin{tabular}{l c c}
		\hline
		Parameter & Setting\\
		\hline
		Coil & Medrad eCoil\\
		TR& 10,000 ms\\
		TE & 73.6 ms\\
		Resolution & $0.55\times 0.55 \times 3$ $mm^{3}$\\\
		Field Of View(FOV) & $14 \times 14$cm\\
		b-value & 100, 400, 1000 $s/mm^{2}$\\
		\hline
	\end{tabular}
\vspace{-0.15in}
\end{table}

The performance of the proposed NCBC method was evaluated using three different experiments: i) synthetic DW-MR phantom acquired using endorectal coil (Experiment 1), ii) physical DW-MR phantom acquired using endorectal coil (Experiment 2), and iii) real patient DW-MR data with endorectal coil at different b-values (Experiment 3). Different quantitative metrics were used to evaluate the results of each designed experiment. In all experiments, the performance of the proposed NCBC method was compared to the following bias correction algorithms: (1) level set method (LS)~\cite{li2011level}, low entropy minimization with a bicubic spline (LEMS)~\cite{salvado2006method}, surface coil intensity correction (SCIC)~\cite{avinash1999method}, spatial gradient distribution (SGD)~\cite{zheng2009automatic}, bias corrected fuzzy C-means (BCFCM)~\cite{ahmed2002modified} and Monte Carlo bias correction (MCBC)~\cite{lui2014monte}. This set of methods comprises of the most widely-used bias correction methods in the literature. All the above algorithms were tested using the provided code by the author of corresponding papers, implemented in MATLAB. The optimal values for the parameters of each algorithm were chosen.
\vspace{-0.1in}
\subsection{Experiment 1: Synthetic Phantom}
\begin{figure}[ht]
	\center
	\subfloat{\includegraphics[trim = 0cm 2.9cm 0cm 0cm, clip,width=1.4in]{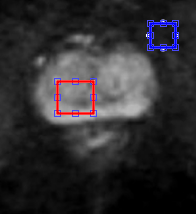}}
	\hspace{1mm}
	\subfloat{\includegraphics[trim = 0cm 2.9cm 0cm 0cm, clip,width=1.4in]{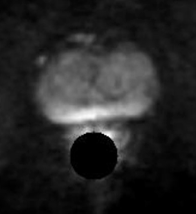}}
	\vspace{-0.1in}
	\caption{Nonendorectal DW-MR data (left) used to generate the synthetic phantom image (right). The red box represents a chosen ROI on the prostate gland and the blue box represents a chosen ROI on the background region. The ROI in red box is used for the SNR calculation. The CNR value is calculated between the red and blue ROIs.}
	\label{fig1}
	\vspace{-0.1in}
\end{figure}

A synthetic DW-MR phantom as shown in Fig.~\ref{fig1} (right) was generated using the same procedure described in~\cite{lui2014monte}, where an artificial bias field with a Gaussian intensity decay is introduced into a nonendorectal DW-MR data sample as shown in Fig.~\ref{fig1} (left).

The proposed NCBC method was applied to reconstruct a noise-free, bias-free endorectal DW-MR image from the generated synthetic phantom acquisition. The effectiveness of the proposed NCBC method on this synthetic phantom test was evaluated using the calculation of correlation coefficient ($r$) as a similarity metric, signal-to-noise ratio (SNR), as well as contrast-to-noise ratio (CNR). The correlation coefficient ($r$) was formulated as
\begin{equation}
r=\frac{\sum_{i}\sum_{j}(V_{ij}-\hat{V})(M_{ij}-\hat{M})}{\sqrt{(\sum_{i}\sum_{j}(V_{ij}-\hat{V})^{2})(\sum_{i}\sum_{j}(M_{ij}-\hat{M})^{2})}}
\end{equation}
\noindent where $\hat{V}$ and $\hat{M}$ correspondingly denote the mean values of uncorrupted ground truth phantom image $V$ as shown in Fig.~\ref{fig1} (left) and reconstructed noise-free, bias-free phantom image $M$. Higher correlation coefficient value $(r=+1)$ means better spatial similarity between the ground-truth in Fig.~\ref{fig1} (left) and the reconstructed synthetic phantom image. To evaluate the performance of the proposed NCBC method in terms of noise suppression, SNR and CNR metrics were calculated as
\begin{equation}
SNR=20\log\frac{\hat{x}_{p}}{\sigma_{p}}
\label{snr}
\end{equation}

\begin{equation}
CNR=20\log\frac{\left|\hat{x}_{b}-\hat{x_{p}}\right|}{\sigma_{b}}.
\label{cnr}
\end{equation}

In the SNR formulation, $\hat{x}_{p}$ represents the mean of a chosen specified Region Of Interest (ROI) on the homogeneous area of prostate gland and $\sigma_p$ denotes the standard deviation of that ROI. In CNR formulation, $\hat{x}_{b}$ and $\hat{x}_{p}$ correspondingly refer to the mean values of chosen ROIs on the background and prostate regions while $\sigma_{b}$ represents the standard deviation of the background region. Better noise suppression should result in higher SNR and CNR values. The regions that are used for the calculation of SNR and CNR values are shown in Fig.~\ref{fig1}(left). Here, the red box represents a chosen ROI on the prostate gland and the blue box is a chosen ROI on the background region. The SNR value is calculated for the red ROI in Fig.~\ref{fig1}(left). The CNR value is calculated between the chosen red and blue ROIs as shown in Fig.~\ref{fig1}(left).
\vspace{-0.15in}
\subsection{Experiment 2: Physical Phantom}
A tissue equivalent ultrasound prostate training phantom from Computerized Imaging Reference Systems Inc. (CIRCS Model 053) was imaged using the same MR scanner with endorectal MR coil. The acquired DW-MR phantom data was utilized to evaluate the performance of the proposed NCBC method in terms of joint bias correction and noise compensation. To evaluate this in a quantitative manner, the coefficient of variation (CV) was calculated using the following formulation:
\begin{equation}
CV=\frac{\sigma_p}{\hat{x}_{p}}.
\label{CV}
\end{equation}
Here, $\hat{x}_{p}$ represents the mean of a chosen ROI on the homogeneous region of prostate gland and $\sigma_p$ denotes the standard deviation of that ROI. To assess the performance of the proposed NCBC method in terms of data noise compensation, SNR and CNR values were also calculated using the formulation in Eq.~\ref{snr} and Eq.~\ref{cnr}. Better bias correction typically should result in lower CV values (less amount of intensity variation) in homogeneous areas, especially at regions closer to the endorectal coil, while better noise suppression should result in higher SNR and CNR values. All the chosen ROIs that are used for the aim of SNR, CNR as well as CV calculations are shown on Fig.~\ref{fig4}(a). Here, the red, blue and green boxes correspondingly represent the selected ROIs on the different areas of prostate gland. The ROI represented by the red box is used for the aim of SNR calculation. The CNR value is calculated between the ROIs represented by the red box and blue box. The green box represents a homogeneous area on the physical phantom that is used for the aim of CV calculation.
\vspace{-0.15in}
\subsection{Experiment 3: Real DW-MR Data}
To quantitatively analyze the performance of the proposed NCBC method using real patient DW-MR data, the Fisher Criterion (FC) analysis was performed on all real patient DW-MR data, and can be expressed as
\begin{equation}
J=\frac{\left|\hat{x}_{b}-\hat{x_{p}}\right|^{2}}{\sigma_{b}^{2}+\sigma_{p}^{2}}
\label{Fisher}
\end{equation}
\noindent where $J$ is the FC value, $\hat{x}_{b}$ and $\hat{x}_{p}$ are the mean of background and prostate classes and $\sigma_{b}$ and $\sigma_{p}$ refer to the variance of background and prostate classes. FC is a valuable metric that can be used for evaluating the power of the proposed NCBC method in improving the delineation of prostate gland and surrounding tissue in the reconstructed DW-MR image. Higher FC values means better separation between the prostate gland and surrounding tissue which leads to the easier and more accurate prostate segmentation and tumor localization.

Another metric that is useful for assessing the performance of the proposed NCBC method in better delineation of prostate gland and surrounding tissue is the probability of error,
\begin{equation}
P(e)=\int min\left[P(\varrho|l),P(\beta|l)\right]P(l)dl.
\label{Perror}
\end{equation}
that is calculated by learning a Bayes classifier~\cite{rish2001empirical} which classifies each DW-MR data to the prostate ($\varrho$) and background ($\beta$) classes and by assuming a normal distribution for each class of prostate and background. Here, $l$ denotes to the DW-MR data at each pixel. The classifier parameters were learned in a Maximum Likelihood framework. Lower probability of error indicates less overlap between the distribution of prostate and background classes which results in better delineation of prostate gland from surrounding tissue in reconstructed noise-free, bias-free endorectal DW-MR image.

To assess the effectiveness of the proposed NCBC method in reconstructed noise-free, bias-free endorectal DW-MR data, SNR (Eq.~\ref{snr}) and CNR (Eq.~\ref{cnr}) values were calculated for all patient cases. To determine the statistical significance of the results, p-value analysis was performed for all these metrics.

\section{Experimental results and Discussion}\label{Results}
In this section, the quantitative results pertaining to each of the three experiments described in Section~\ref{sec:exp} are presented and discussed for assessing the efficacy of the proposed NCBC reconstruction method when compared to existing methods. Furthermore, visual results produced using the tested methods corresponding to each of the three experiments are presented and discussed to provide a qualitative assessment of the proposed method. Finally, a computation time analysis is performed to assess the computational efficiency of the proposed method compared to existing methods.

\begin{figure}[!hpt]
	\vspace{-0.2in}
	\centering
	\subfloat[Ground truth]{\includegraphics[trim = 1cm 2.9cm 1cm 1cm, clip,width=0.85in]{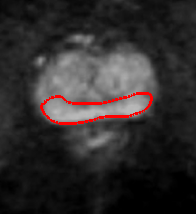}}
	\subfloat[UC]{\includegraphics[trim = 1cm 2.9cm 1cm 1cm, clip,width=0.85in]{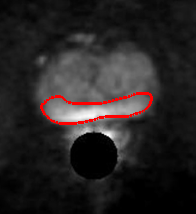}}
	\subfloat[LS]{\includegraphics[trim = 1cm 2.9cm 1cm 1cm, clip,width=0.85in]{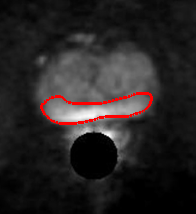}}
	\vspace{-0.14in}	
	\subfloat[LEMS]{\includegraphics[trim = 1cm 2.9cm 1cm 1cm, clip,width=0.85in]{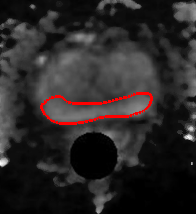}}
	\subfloat[SCIC]{\includegraphics[trim = 1cm 2.9cm 1cm 1cm, clip,width=0.85in]{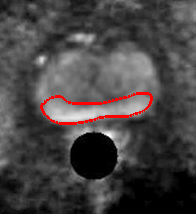}}
	\subfloat[SGD]{\includegraphics[trim = 1cm 2.9cm 1cm 1cm, clip,width=0.85in]{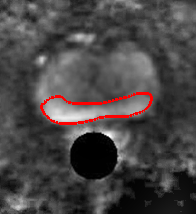}}
	\vspace{-0.14in}
	\subfloat[BCFCM]{\includegraphics[trim = 1cm 2.9cm 1cm 1cm, clip,width=0.85in]{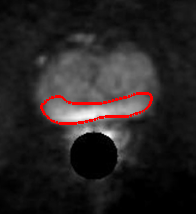}}
	\subfloat[MCBC]{\includegraphics[trim = 1cm 2.9cm 1cm 1cm, clip,width=0.85in]{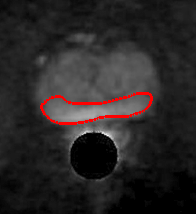}}
	\subfloat[NCBC]{\includegraphics[trim = 1cm 2.9cm 1cm 1cm, clip,width=0.85in]{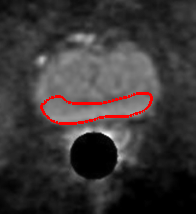}}
	\caption{(a) Nonendorectal DW-MR data (ground truth). (b-i) Reconstructed synthetic DW-phantom images using different tested methods. NCBC shows better bias correction compared to the UC-phantom as well as all other bias correction methods. The red ROIs show a region of peripheral zone (PZ) of the prostate gland that bias field present to a significant degree in reconstructed images using LS, SCIC, SGD and BCFCM methods, while strong inhomogeneity correction performance is achieved using MCBC, LEMS, as well as NCBC methods.}
	\label{fig2}
\end{figure}
\vspace{-0.5em}

\begin{figure}[!hpt]
	\vspace{-0.3in}
	\centering
	\subfloat[]{\includegraphics[trim = 1cm 2.9cm 1cm 1cm, clip,height=0.67in,width=0.9in]{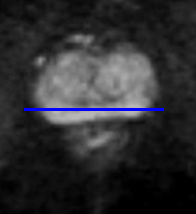}}
	\hspace{2.5mm}
	\subfloat[]{\includegraphics[trim = 0cm 0cm 0cm 0cm, clip,height=0.73in,width=1.1in]{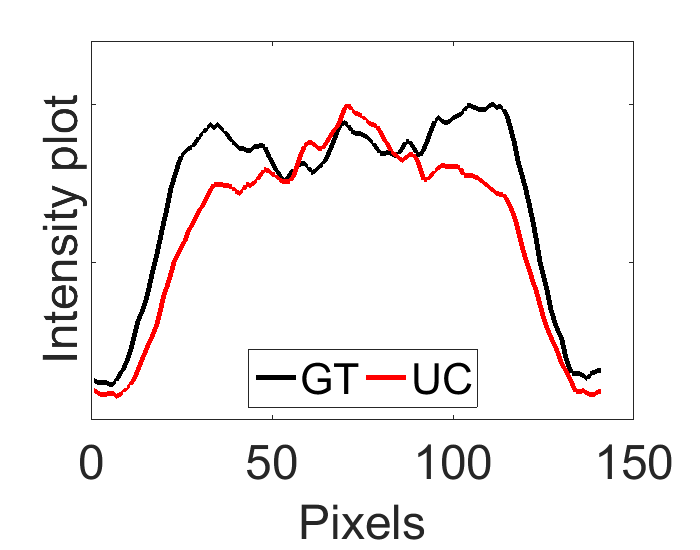}}
	\subfloat[]{\includegraphics[trim = 0cm 0cm 0cm 0cm, clip,height=0.73in,width=1.1in]{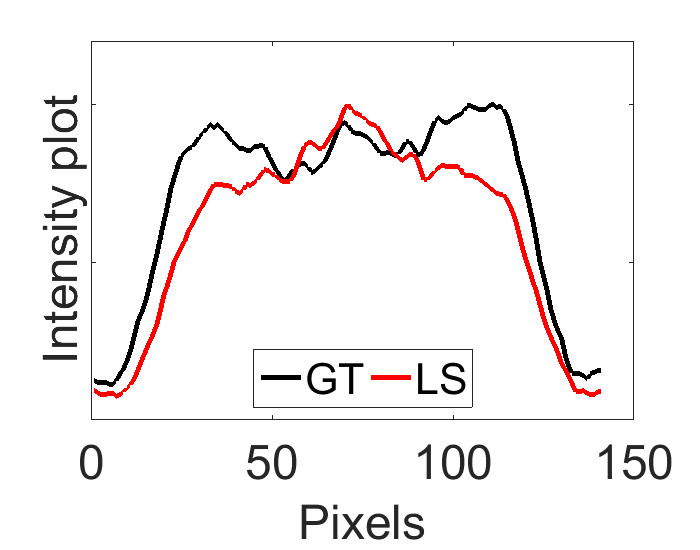}}
	\vspace{-0.14in}
	\subfloat[]{\includegraphics[trim = 0cm 0cm 0cm 0cm,clip,height=0.73in,width=1.1in]{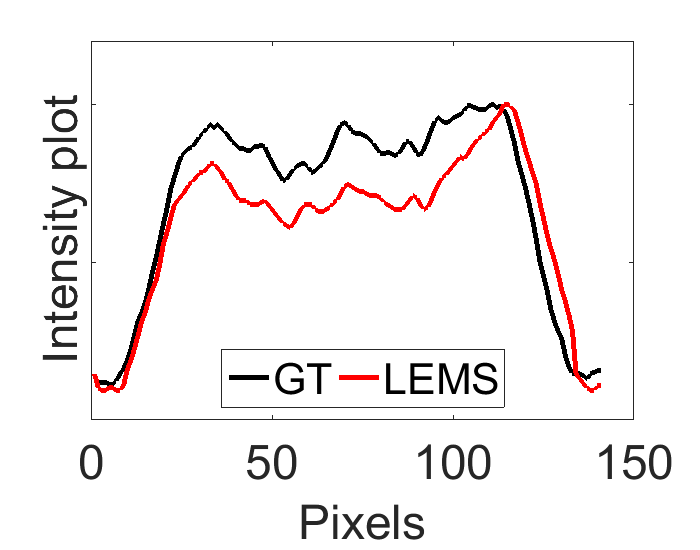}}
	\subfloat[]{\includegraphics[trim = 0cm 0cm 0cm 0cm,clip,height=0.73in,width=1.1in]{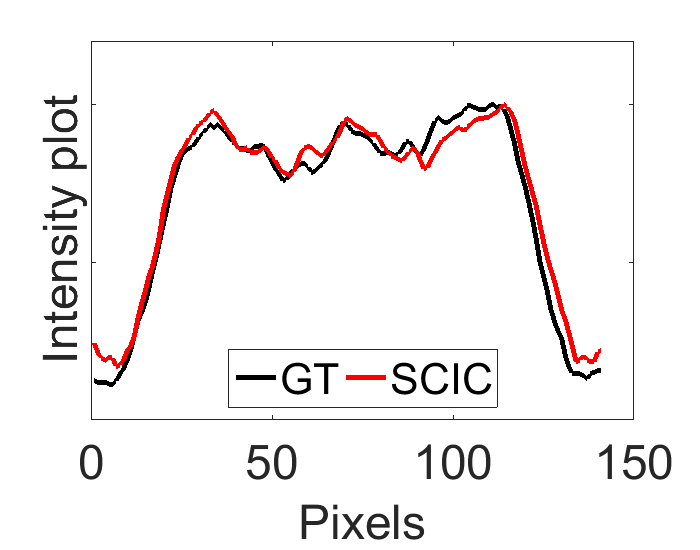}}
	\subfloat[]{\includegraphics[trim = 0cm 0cm 0cm 0cm, clip,height=0.73in,width=1.1in]{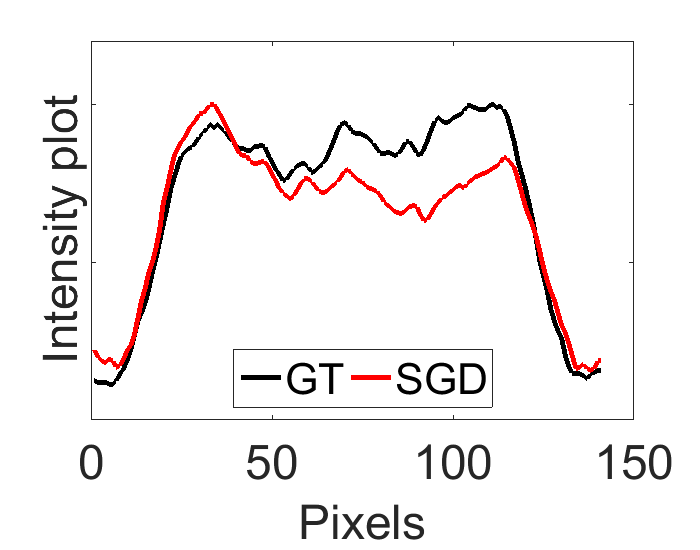}}
	\vspace{-0.14in}
	\subfloat[]{\includegraphics[trim = 0cm 0cm 0cm 0cm, clip,height=0.73in,width=1.1in]{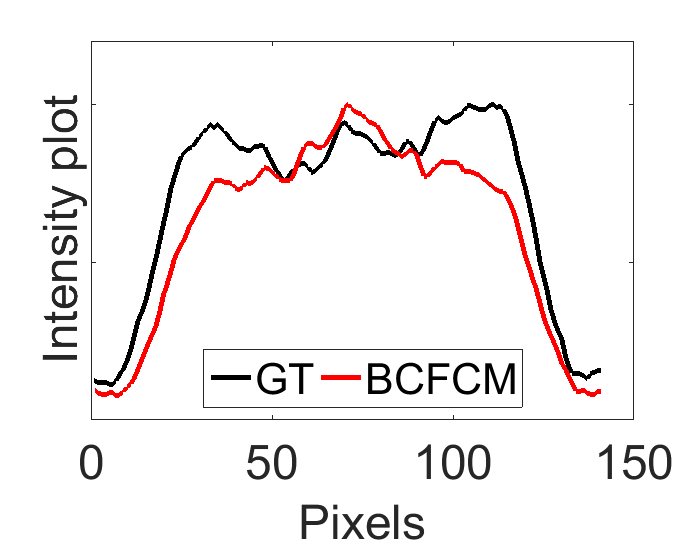}}
	\subfloat[]{\includegraphics[trim = 0cm 0cm 0cm 0cm,clip,height=0.73in,width=1.1in]{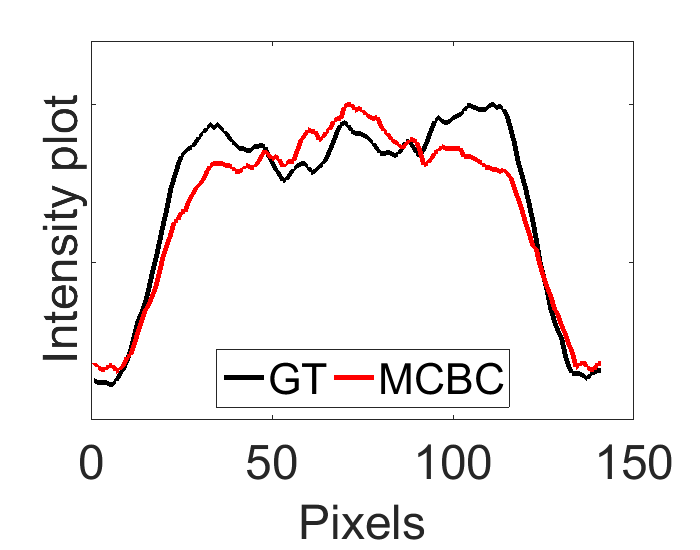}}
	\subfloat[]{\includegraphics[trim = 0cm 0cm 0cm 0cm,clip,height=0.73in,width=1.1in]{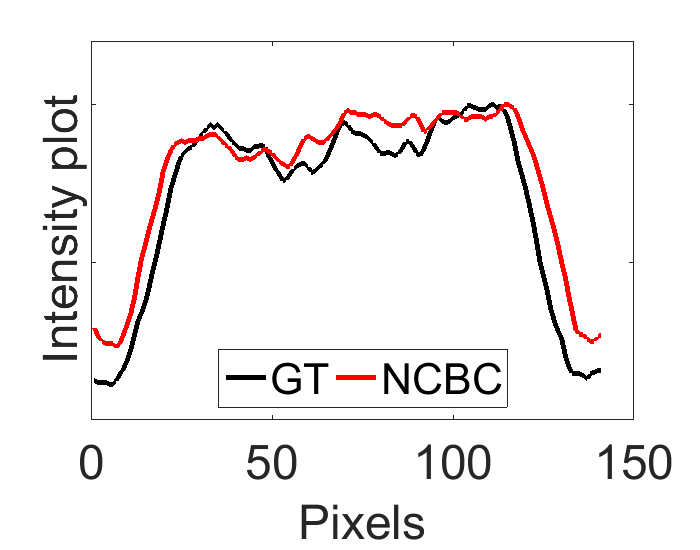}}
	\caption{(a) Nonendorectal DW-MR data (ground truth). The data intensity profiles corresponding to the blue line in (a) are shown in (b) for uncorrected data (red color plot) versus the ground truth (GT) data (black color plot) and and in (c-i) for the corrected data using different methods (red color plots) versus the ground truth data (black color plots). NCBC shows the most flat intensity profile with the minimal intensity variation compared to the other tested methods.}
	\label{fig3}
\end{figure}
\vspace{-1em}
\subsection{Experiment 1: Synthetic Phantom}
The calculated correlation coefficient ($r$), SNR, as well as CNR values for all tested methods in the synthetic phantom experiment are shown in Table~\ref{Ta2}. As seen in Table~\ref {Ta2}, the proposed NCBC method achieved the highest $r$, CNR as well as SNR values when compared to that of other tested methods. Visual results from the bias-corrected DW-MR synthetic phantom data produced using different tested methods is shown in Fig.~\ref{fig2}(c-i). It can be observed that the MCBC, LEMS, and proposed NCBC method were able to achieve the best level of bias correction when compared to the other tested methods. This is particularly apparent in the peripheral zone (PZ) of the prostate gland as highlighted using red ROIs in Fig.~\ref{fig2}(a-i), where intensity inhomogeneities are still present to a significant degree in the bias-corrected phantom data produced using the LS, SCIC, SGD and BCFCM methods, while strong inhomogeneity correction performance is achieved using MCBC, LEMS, and the proposed NCBC method. Furthermore, it can be observed that NCBC exhibited minimal intensity inhomogeneities when compared with MCBC and LEMS, particularly in the area that is highlighted using Red ROIs. To better represent the outperforming of proposed NCBC method in terms of bias correction using synthetic phantom, the data intensity profiles corresponding to the drawn blue line in Fig.~\ref{fig3}(a) are shown in Fig.~\ref{fig3}(b) for uncorrected data (red color plot) versus ground truth data (black color plot) and in Fig.~\ref{fig3}(c-i) for corrected data using different tested methods (red color plots) versus ground truth data (black color plots). As the intensity profiles of Fig.~\ref{fig3}(b-i) show, NCBC method was resulted in an intensity profile with the most flatness and less amount of variation compared to the intensity profiles of uncorrected image as well as reconstructed images using other tested methods as such confirms the better performance of proposed NCBC method in terms of bias field correction.

\begin{table}[!h]
	\scriptsize
	\renewcommand{\arraystretch}{1.3}
	\caption{correlation coefficient, cnr (db) and snr (db) for synthetic phantom experiment. boldface indicates highest performance.}
	\label{Ta2}
	\centering
	\begin{tabular}{|c||c||c||c|}
		\hline
		\bf Method & \bf Correlation Coefficient (r) &\bf CNR (dB) &\bf SNR (dB)\\
		\hline
		\hline
		\bf UC & 0.8291 &  20.26 & 20.15\\
		\hline
		\bf SCIC & 0.8632 & 20.68 &  20.76\\
		\hline
		\bf SGD & 0.8620 & 17.77 &  19.46\\
		\hline
		\bf LS & 0.8907 & 20.26 & 20.25\\
		\hline
		\bf LEMS & 0.8698 & 21.05 & 22.90\\
		\hline
		\bf BCFCM & 0.8521& 20.32 & 20.25\\
		\hline
		\bf MCBC & 0.9484 & 22.49 & 23.28\\
		\hline
		\bf NCBC & \bf 0.9815 &\bf23.48 & \bf25.17\\
		\hline
	\end{tabular}
\vspace{-0.3in}
\end{table}

\begin{figure}[!hpt]
	\centering
	\subfloat[UC]{\includegraphics[trim = 1.5cm 2cm 1.5cm 0.7cm, clip,width=0.85in]{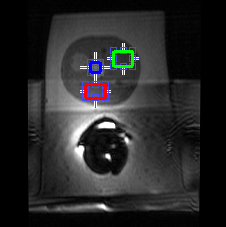}}
	\subfloat[LS]{\includegraphics[trim = 1.5cm 2cm 1.5cm 0.7cm, clip,width=0.85in]{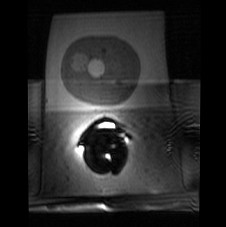}}
	\subfloat[LEMS]{\includegraphics[trim = 1.5cm 2cm 1.5cm 0.7cm, clip,width=0.85in]{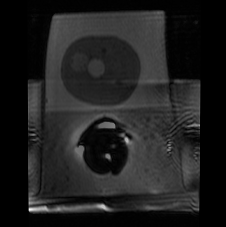}}
	\subfloat[SCIC]{\includegraphics[trim = 1.5cm 2cm 1.5cm 0.7cm, clip,width=0.85in]{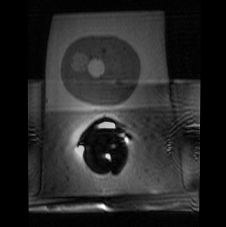}}
\vspace{-0.14in}	
	\subfloat[SGD]{\includegraphics[trim = 1.5cm 2cm 1.5cm 0.7cm, clip,width=0.85in]{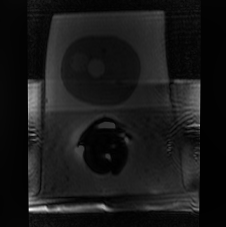}}
	\subfloat[BCFCM]{\includegraphics[trim = 1.5cm 2cm 1.5cm 0.7cm, clip,width=0.85in]{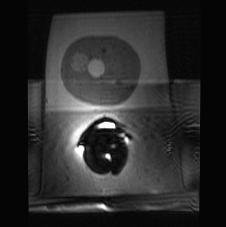}}
	\subfloat[MCBC]{\includegraphics[trim = 1.5cm 2cm 1.5cm 0.7cm, clip,width=0.85in]{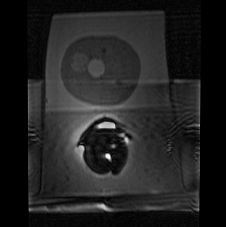}}
	\subfloat[NCBC]{\includegraphics[trim = 1.5cm 2cm 1.5cm 0.7cm, clip,width=0.85in]{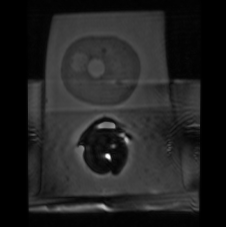}}
	\caption{NCBC method shows strong bias field correction when compared to the UC physical phantom image as well as reconstructed images using other tested bias correction methods. Red, blue and green boxes represent different selected ROIs on the prostate gland. The SNR calculation is performed for the ROI inside of the red box. The CNR value is calculated between the ROIs represented by the red box and blue box. Green box represents a selected homogeneous area of physical phantom that is used for CV calculation.
			}
\vspace{-0.2in}
	\label{fig4}
\end{figure}
\subsection{Experiment 2: Physical Phantom}
The calculated SNR, CNR, and CV values for the tested methods in the physical phantom experiment are shown in Table~\ref{Ta3}. As seen in Table~\ref {Ta3}, the proposed NCBC method achieved the highest SNR and CNR values, and the lowest CV values (lower indicates better bias correction performance), when compared to that of other tested methods. Visual results from the bias-corrected DW-MR physical phantom data produced using the tested methods is shown in Fig.~\ref{fig4}(b-h). Similar to the results of the synthetic phantom experiment, it can be observed that NCBC exhibited minimal intensity inhomogeneities when compared with all other tested methods, which is particularly noticeable in the regions surrounding the endorectal coil. Furthermore, it can also be observed that the reconstructed DW-MR data produced using NCBC also exhibit less data noise when compared to the other tested methods. As such, based on visual assessment, it can be clearly seen from the phantom experiments so far that the proposed NCBC method achieves improved bias correction performance when compared to the other tested methods and is capable to suppressing data noise (which usually becomes amplified as a result of bias correction) by accounting for both bias and data noise within an unified framework.

\begin{table}[!h]
	\scriptsize
	\renewcommand{\arraystretch}{1.3}
	\caption{snr (db), cnr (db), and cv values for physical phantom experiment. boldface indicates highest performance (lower is better for cv).}
	\label{Ta3}
	\centering
	\begin{tabular}{|c||c||c||c|}
		\hline
		\bf Method & \bf SNR (dB) & \bf CNR (dB) & \bf CV\\
		\hline
		\hline
		\bf UC &  22.596 &  5.910 & 0.182\\
		\hline
		\bf SCIC &  25.131 & 13.961  & 0.108\\
		\hline
		\bf SGD &   22.133 & 11.183 & 0.123\\
		\hline
		\bf LS &    22.596 & 5.910  &   0.154\\
		\hline
		\bf LEMS &  25.158 & 10.717 & 0.085\\
		\hline
		\bf BCFCM &  23.418 & 8.391 &   0.136\\
		\hline
		\bf MCBC &   25.508 & 12.891 & 0.106\\
		\hline
		\bf NCBC & \bf32.899 & \bf15.768 & \bf 0.039\\
		\hline
	\end{tabular}
\vspace{-0.2in}
\end{table}

\subsection{Experiment 3: Real DW-MR Data}
To assess the effectiveness of the proposed NCBC method in a real-world clinical scenario, DW-MR data from 7 patient cases, each consisting of two acquisitions with different b-value ($b=100 s/mm^{2}$ and $b=1000 s/mm^{2}$), was used in this experiment. Different performance metrics including SNR, CNR, FC, and probability of error were used for quantitative assessment. The quantitative results, along with visual assessment of the results, are provided below.
\subsubsection{SNR and CNR Analysis}

The calculated SNR values for the tested methods for both b-value acquisitions in the real DW-MR data experiment are shown in Table~\ref{Ta4}. As seen in Table~\ref {Ta4}, the proposed NCBC method achieved the highest SNR values for all acquisitions except one case and when compared to that of other tested methods, with an average improvement of 6$dB$ over the uncorrected DW-MR data. Likewise, the calculated CNR values for the tested methods for both b-value acquisitions in the real DW-MR data experiment are shown in Table~\ref{Ta5}. As seen in Table~\ref {Ta5}, the proposed NCBC method also achieved the highest CNR values in most cases when compared to that of other tested methods, with an average improvement of 5.1$dB.$ over the uncorrected DW-MR data.
To analyze the statistical significance of the SNR and CNR values, p-values were calculated and reported in Table \ref{Ta9}. In the p-value analysis, the null assumption was that a given reconstruction method had no improvement over the SNR or CNR values as compared to the uncorrected DW-MR data. All p-values are computed using a two-tailed normal distribution with a statistical significance level of $0.05$. The p-values for the SNR analysis demonstrate the statistical significance of all methods in SNR improvement, with the lowest p-value obtained for the proposed NCBC method. Furthermore, the calculated p-values based on the CNR analysis shows that the proposed NCBC method had the lowest p-value among all tested methods and is demonstrated to be statistical significant. The CNR p-value results also show the statistical significance of all tested methods except SGD and LS methods which resulted in p-values more than $0.05$. To evaluate the effectiveness of proposed NCBC reconstruction method in terms of improving the visualization of prostate tumor, CNR values between the tumor and normal tissue is also calculated using ADC map images for the 2 patients with prostate tumor. Calculated CNR values are reported in Table~\ref{Ta6} when the proposed NCBC method achieved the highest CNR values for both patient cases and compared to that of other tested methods as well as UC image.

\begin{figure}[!hpt]
	\centering
	\subfloat[UC]{\includegraphics[trim = 2cm 3cm 2cm 3.5cm, clip,width=0.85in]{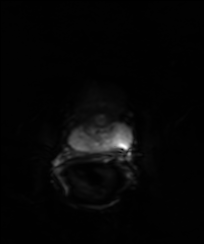}}
	\subfloat[LS]{\includegraphics[trim = 2cm 3cm 2cm 3.5cm, clip,width=0.85in]{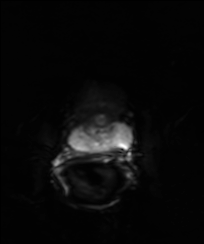}}
	\subfloat[LEMS]{\includegraphics[trim = 2cm 3cm 2cm 3.5cm, clip,width=0.85in]{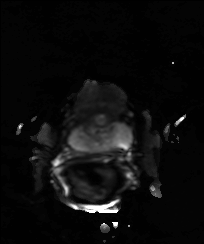}}
	\subfloat[SCIC]{\includegraphics[trim =  2cm 3cm 2cm 3.5cm, clip,width=0.85in]{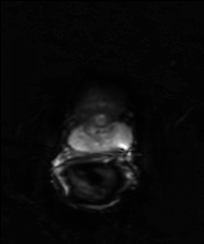}}
	\vspace{-0.14in}
	\subfloat[SGD]{\includegraphics[trim =  2cm 3cm 2cm 3.5cm, clip,width=0.85in]{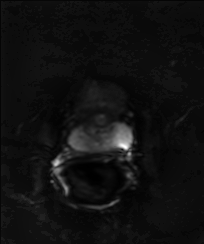}}
	\subfloat[BCFCM]{\includegraphics[trim = 2cm 3cm 2cm 3.5cm, clip,width=0.85in]{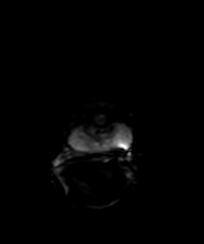}}
	\subfloat[MCBC]{\includegraphics[trim = 2cm 3cm 2cm 3.5cm, clip,width=0.85in]{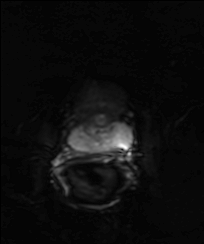}}
	\subfloat[NCBC]{\includegraphics[trim =  2cm 3cm 2cm 3.5cm, clip,width=0.85in]{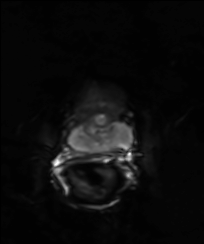}}
	\caption{Case 1 with b=100 $s/mm^{2}$. NCBC method reduces the bias field and noise in prostate DWI.}
	\label{fig5}
\vspace{-0.4in}
\end{figure}
\begin{figure}[!hpt]
	\centering
	\subfloat[UC]{\includegraphics[trim = 1.8cm 3.2cm 1.8cm 3.5cm, clip,width=0.85in]{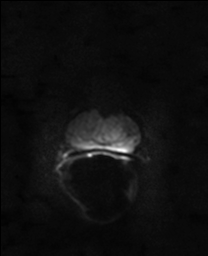}}
	\subfloat[LS]{\includegraphics[trim = 1.8cm 3.2cm 1.8cm 3.5cm, clip,width=0.85in]{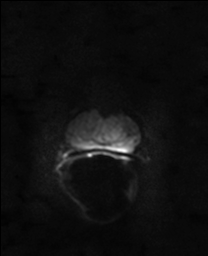}}
	\subfloat[LEMS]{\includegraphics[trim = 1.8cm 3.2cm 1.8cm 3.5cm, clip,width=0.85in]{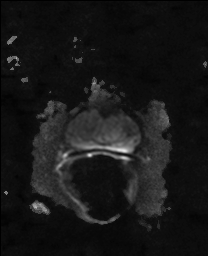}}
	\subfloat[SCIC]{\includegraphics[trim = 1.8cm 3.2cm 1.8cm 3.5cm, clip,width=0.85in]{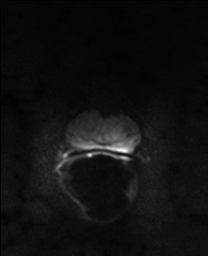}}
	\vspace{-0.14in}
	\subfloat[SGD]{\includegraphics[trim = 1.8cm 3.2cm 1.8cm 3.5cm, clip,width=0.85in]{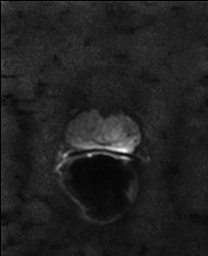}}
	\subfloat[BCFCM]{\includegraphics[trim = 1.8cm 3.2cm 1.8cm 3.5cm, clip,width=0.85in]{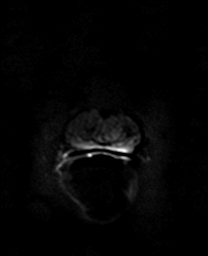}}
	\subfloat[MCBC]{\includegraphics[trim = 1.8cm 3.2cm 1.8cm 3.5cm, clip,width=0.85in]{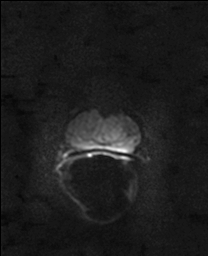}}
	\subfloat[NCBC]{\includegraphics[trim = 1.8cm 3.2cm 1.8cm 3.5cm, clip,width=0.85in]{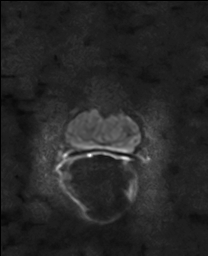}}
	\caption{Case 2 with b=1000 $s/mm^{2}$. NCBC method reduces the bias field and noise in prostate DWI.}
	\label{fig6}
	\vspace{-0.1in}
\end{figure}
\begin{figure}[!hpt]
	\centering
	\subfloat[UC]{\includegraphics[trim = 0cm 1cm 0cm 0.3cm, clip,width=0.85in]{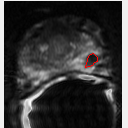}}
	\subfloat[LS]{\includegraphics[trim = 0cm 1cm 0cm 0.3cm, clip,width=0.85in]{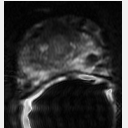}}
	\subfloat[LEMS]{\includegraphics[trim = 0cm 1cm 0cm 0.3cm, clip,width=0.85in]{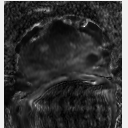}}
	\subfloat[SCIC]{\includegraphics[trim =0cm 1cm 0cm 0.3cm, clip,width=0.85in]{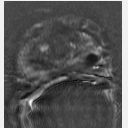}}
	\vspace{-0.14in}
	\subfloat[SGD]{\includegraphics[trim =0cm 1cm 0cm 0.3cm, clip,width=0.85in]{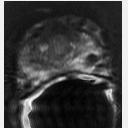}}
	\subfloat[BCFCM]{\includegraphics[trim =0cm 1cm 0cm 0.3cm, clip,width=0.85in]{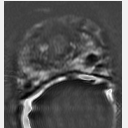}}
	\subfloat[MCBC]{\includegraphics[trim =0cm 1cm 0cm 0.3cm, clip,width=0.85in]{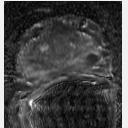}}
	\subfloat[NCBC]{\includegraphics[trim =0cm 1cm 0cm 0.3cm, clip,width=0.85in]{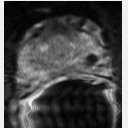}}
	\caption{ADC images for a patient with tumor reconstructed using different tested methods. The tumor location is marked using red contour in (a). NCBC method reduces the bias field and noise in prostate DWI such that provides better representation of prostate tumor.}
	\label{fig7}
\end{figure}

\begin{figure}[!hpt]
	\centering
	\subfloat[UC]{\includegraphics[trim = 0.25cm 1.5cm 0.3cm 0.4cm, clip,width=0.85in]{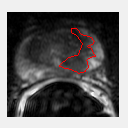}}
	\subfloat[LS]{\includegraphics[trim = 0.25cm 1.5cm 0.3cm 0.4cm, clip,width=0.85in]{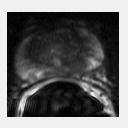}}
	\subfloat[LEMS]{\includegraphics[trim =0.25cm 1.5cm 0.3cm 0.4cm, clip,width=0.85in]{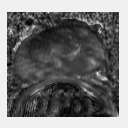}}
	\subfloat[SCIC]{\includegraphics[trim =0.25cm 1.5cm 0.3cm 0.4cm, clip,width=0.85in]{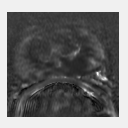}}
	\vspace{-0.14in}
	\subfloat[SGD]{\includegraphics[trim =0.25cm 1.5cm 0.3cm 0.4cm, clip,width=0.85in]{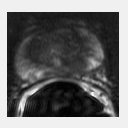}}
	\subfloat[BCFCM]{\includegraphics[trim =0.25cm 1.5cm 0.3cm 0.4cm, clip,width=0.85in]{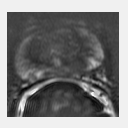}}
	\subfloat[MCBC]{\includegraphics[trim =0.25cm 1.5cm 0.3cm 0.4cm, clip,width=0.85in]{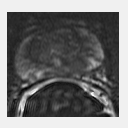}}
	\subfloat[NCBC]{\includegraphics[trim = 0.25cm 1.5cm 0.3cm 0.4cm, clip,width=0.85in]{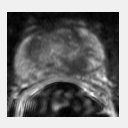}}
	\caption{ADC images for a patient with tumor reconstructed using different tested methods. The tumor location is marked using red contour in (a). NCBC method reduces the bias field and noise in prostate DWI such that provides better representation of prostate tumor.}
	\label{fig8}
	\vspace{-0.2in}
	\end{figure}
	
\begin{figure}[!hpt]
	\centering
	\subfloat[UC]{\includegraphics[trim =  0.25cm 1.5cm 0.3cm 0.4cm, clip,width=1.5in]{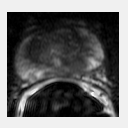}}
	\subfloat[NCBC]{\includegraphics[trim =  0.25cm 1.5cm 0.3cm 0.4cm, clip,width=1.5in]{Fig/Boroo60}}
	\caption{Enlarged ADC images for the patient in Fig.~\ref{fig8} with tumor. (a) Uncorrected image, (b) reconstructed image using proposed NCBC method.}
	\label{fig9}
	\vspace{-0.1in}
	\end{figure}	
\begin{table}[!h]
	\renewcommand{\arraystretch}{1.3}
	\caption{snr (db) values for real dw-mr data experiment. boldface indicates highest performance.}
	\label{Ta4}
	\centering
	\tabcolsep 2.5pt
	\scriptsize
	\begin{tabular}{ccccccccc}
		\hline
		\bf Case & \bf UC & \bf SCIC & \bf SGD & \bf LS & \bf LEMS & \bf BCFCM & \bf MCBC& \bf NCBC\\
		\hline
		1 & 14.7 & 17.0 & 19.3 & 15.3 & 16.0 & 13.6 & 21.0 & \bf22.0 \\
		\hline
		2 & 16.7 & 17.1 & 16.6 & 16.6 & 18.2 & 15.4 & 17.5 & \bf21.1 \\
		\hline
		3 & 16.2 & 17.7 & 14.9 & 16.8 & 18.8 & 13.8 & 20.6 & \bf21.8 \\
		\hline
		4 & 16.3 & 16.5 & 15.9 & 16.4 & 16.6 & 15.4 & 17.6 & \bf17.8 \\
		\hline
		5 & 11.7 & 12.0 & 12.3 & 12.0 & 12.1 & 9.9 & 14.1 & \bf17.6 \\
		\hline
		6 & 8.9 & 8.7 & 8.3 & 8.7 & 9.3 & 5.0 & 11.1 & \bf13.6 \\
		\hline
		7 & 13.6 & 16.7 & 18.0 & 14.2 & 16.1 & 12.5 & 19.8 & \bf20.8 \\
		\hline
		8 & 11.1 & 14.1 & 13.7 & 11.6 & 15.1 & 10.2 & 16.9 & \bf17.9 \\
		\hline
		9 & 12.4 & 14.1 & 13.4 & 12.4 & 15.3 & 11.0 & 17.0 & \bf19.9 \\
		\hline
		10 & 17.1 & 18.6 & 18.9 & 17.5 & 20.3 & 15.5 & \bf21.9 & 21.3 \\
		\hline
		11 & 12.5 & 14.4 & 12.9 & 12.9 & 15.8 & 6.9 & 16.3 & \bf18.7 \\
		\hline
		12 & 7.6 & 8.0 & 7.8 & 7.7 & 8.2 & 3.8 & 9.8 & \bf13.8 \\
		\hline
		13 & 15.1 & 19.4 & 15.9 & 15.1 & 20.8 & 11.1 & 19.1 & \bf25.8 \\
		\hline
		14 & 13.4 & 15.0 & 15.2 & 13.7 & 15.3 & 8.5 & 18.4 & \bf19.0 \\
		\hline
		Ave. & 13.4 & 15.0 & 14.5 & 13.6 & 15.6 & 10.9 & 17.2 & \bf19.4 \\
		\hline
		\end{tabular}
\end{table}

\begin{table}[!htp]
	\renewcommand{\arraystretch}{1.3}
	
	\caption{cnr (dB) values for real dw-mr data experiment. boldface indicates highest performance.}
	\label{Ta5}
	\centering
	\tabcolsep 2.5pt
	\scriptsize
	\begin{tabular}{ccccccccc}
		\hline
		\bf Case & \bf UC & \bf SCIC & \bf LS & \bf LEMS & \bf SGD & \bf BCFCM & \bf MCBC& \bf NCBC\\
		\hline
		1 & 11.5 & 12.5 & 13.1 & 11.8 & 11.2 & 11.0 & 15.0 & \bf15.8 \\
		\hline
		2 & 12.1 & 12.3 & 12.1 & 12.2 & 10.9 & 11.9 & 12.5 & \bf17.3 \\
		\hline
		3 & 15.8 & 16.9 & 11.1 & 16.1 & 4.5 & 15.3 & 18.5 & \bf20.7 \\
		\hline
		4 & 21.0 & 21.6 & 21.1 & 21.4 & 21.2 & 21.1 & 21.2 & \bf24.3 \\
		\hline
		5 & 11.7 & 13.2 & 11.5 & 11.9 & 5.0 & 11.3 & 14.1 & \bf20.6 \\
		\hline
		6 & 13.0 & 13.4 & 11.7 & 13.1 & 12.6 & 11.5 & 11.8 &\bf 16.1 \\
		\hline
		7 & 13.2 & 13.9 & 14.9 & 13.4 & 12.9 & 12.6 & 17.5 & \bf17.9 \\
		\hline
		8 & 17.2 & \bf21.0 & 19.7 & 17.6 & 16.7 & 17.9 & 18.3 & 18.2 \\
		\hline
		9 & 15.9 & 18.9 & 16.9 & 15.9 & 19.9 & 16.5 & 16.0 & \bf24.7 \\
		\hline
		10 & 23.5 & 25.3 & 24.8 & 23.6 & 26.9 & 22.8 & \bf26.1 & 24.4 \\
		\hline
		11 & 14.0 & 16.1 & 13.3 & 14.4 & 17.2 & 10.4 & 17.0 & \bf19.9 \\
		\hline
		12 & 12.1 & 12.5 & 11.6 & 12.1 & 12.2 & 10.4 & 14.5 & \bf23.8 \\
		\hline
		13 & 12.2 & 13.2 & 12.0 & 12.4 & 13.0 & 9.2 & 15.3 & \bf16.7 \\
		\hline
		14 & 15.7 & 17.3 & 17.6 & 15.8 & 16.8 & 13.0 & 19.1 &\bf 19.3 \\
		\hline
		Ave. & 14.9 & 16.3 & 15.1 & 15.1 & 14.4 & 13.9 & 16.9 & \bf20.0 \\
       \hline
     \end{tabular}
		\end{table}
\begin{table}[!htp]
	\renewcommand{\arraystretch}{1.3}
	\caption{cnr (dB) between tumor and normal tissue calculated using adc images. boldface indicates highest performance.}
	\label{Ta6}
	\centering
	\tabcolsep 2.5pt
	\scriptsize
	\begin{tabular}{ccccccccc}
		\hline
		\bf Case & \bf UC & \bf SCIC & \bf LS & \bf LEMS & \bf SGD & \bf BCFCM & \bf MCBC& \bf NCBC\\
		\hline
		1 & 22.5 & 18.2 & 21.8 & 12.8 & 22.7 & 20.1 & 18.2 & \bf24.8 \\
		\hline
		2& 19.6 & 22.2 & 19.6 & 10.4 & 18.7 & 22.0 & 19.2 & \bf27.3 \\
		\hline
	\end{tabular}
\end{table}
\subsubsection{Fisher Criterion}
To assess the performance of proposed NCBC method in terms of achieving separability of the prostate gland from the surrounding tissue, FC analysis (Eq.~\ref{Fisher}) was performed on all 14 acquisitions and for all different tested methods. The FC results are reported in Table~\ref{Ta7} showed that the proposed NCBC method achieved the highest overall FC results, followed by MCBC and SCIC. All other tested methods except BCFCM also demonstrated improvements in FC when compared to the uncorrected DW-MR data, but to a much lesser degree than the proposed NCBC method. The p-value analysis was also performed to evaluate the statistical significance of acquired FC values. The calculated p-values are reported in Table \ref{Ta9}, with the smallest p-value achieved using the proposed NCBC method.

\begin{table}[!h]
	\renewcommand{\arraystretch}{1.3}
    \caption{fisher criterion (fc) values for real dw-mr data experiment. boldface indicates highest performance.}
	\label{Ta7}
	\centering
	\tabcolsep 2.5pt
	\scriptsize
	\begin{tabular}{c c c c c c c c c}
		\hline
		\bf Case & \bf UC & \bf SCIC & \bf SGD & \bf LS & \bf LEMS & \bf BCFCM & \bf MCBC & \bf NCBC\\
		\hline
		1 & 0.94 & 1.29 & 1.68 & 1.00 & 0.66 & 0.98 & 4.25 & \bf7.86 \\
		\hline
		2 & 1.08 & \bf1.23 & 0.95 & 1.10 & 0.94 & 0.79 & 1.03 & 1.05 \\
		\hline
		3 & 0.72 & 1.27 & 0.66 & 0.83 & 1.48 & 0.71 & 3.55 & \bf5.31 \\
		\hline
		4 & 2.29 & 7.80 & 4.18 & 2.76 & 3.22 & 2.20 & 8.36 & \bf10.76 \\
		\hline
		5 & 1.63 & 1.89 & 2.02 & 1.77 & 1.87 & 1.37 & 2.55 & \bf3.89 \\
		\hline
		6 & 1.74 & 4.46 & 1.59 & 2.21 & 2.36 & 1.62 & 3.55 & \bf7.35 \\
		\hline
		7 & 0.81 & 1.07 & 1.32 & 0.86 & 0.91 & 0.78 & 2.74 & \bf5.59 \\
		\hline
		8 & 1.02 & 1.64 & 1.61 & 1.10 & 1.56 & 1.02 & 1.71 & \bf3.77 \\
		\hline
		9 & 0.81 & 1.11 & 1.00 & 0.81 & 2.11 & 0.88 & 1.50 & \bf3.29 \\
		\hline
		10 & 5.65 & 8.06 & 8.47 & 5.65 & 12.20 & 5.90 & 6.64 & \bf16.39 \\
		\hline
		11 & 1.80 & 2.89 & 2.01 & 1.96 & 4.42 & 0.92 & 6.49 & \bf9.87 \\
		\hline
		12 & 1.00 & 1.14 & 1.05 & 1.02 & 1.16 & 0.55 & \bf5.37 & 2.48 \\
		\hline
		13 & 1.02 & 1.29 & 1.13 & 1.08 & 0.99 & 0.50 & \bf4.65 & 3.82 \\
		\hline
		14 & 2.53 & 3.45 & 3.56 & 2.67 & 0.41 & 1.14 & 3.13 & \bf4.02 \\
		\hline
		Ave. & 1.64 & 2.76 & 2.23 & 1.77 & 2.45 & 1.38 & 3.96 & \bf6.10 \\
		\hline
	\end{tabular}
\end{table}

\subsubsection{Probability of Error}
A Bayes classifier was trained to classify all datasets to prostate and background classes, and the probability of error classification was calculated and reported in Table \ref{Ta8}. According to the probability of error results, the proposed NCBC method results in the best separation between the prostate gland from surrounding tissue as the calculated probability of error was considerably lower compared to the other tested methods. These results also demonstrate that the MCBC and SCIC methods were the next two successful methods in terms of prostate classification, while the other tested approaches except BCFCM could also decrease the probability of error classification but to a much lesser extent. In general, these results agree well with the conclusions drawn from the FC analysis.
\begin{table}[!htp]
	\renewcommand{\arraystretch}{1.3}
	
	\caption{probability of error for real dw-mr data experiment. boldface indicates highest performance.}
	\label{Ta8}
	\centering
	\tabcolsep 2.5pt
	\scriptsize
	\begin{tabular}{c c c c c c c c c}
		\hline
		\bf Case & \bf UC & \bf SCIC & \bf SGD & \bf LS & \bf LEMS & \bf BCFCM & \bf MCBC & \bf NCBC\\
		\hline
		1 & 0.090 & 0.049 & 0.065 & 0.079 & 0.065 & 0.146 & 0.024 &  \bf0.011 \\
		\hline
		2 & 0.089 & 0.049 & 0.066 & 0.079 & 0.066 & 0.146 & 0.025 &  \bf0.011 \\
		\hline
		3 & 0.103 & 0.058 & 0.072 & 0.090 & 0.071 & 0.140 & 0.031 &  \bf0.013 \\
		\hline
		4 & 0.089 & 0.048 & 0.063 & 0.077 & 0.063 & 0.141 & 0.025 &  \bf0.012 \\
		\hline
		5 & 0.092 & 0.053 & 0.069 & 0.083 & 0.067 & 0.148 & 0.027 &  \bf0.012 \\
		\hline
		6 & 0.099 & 0.057 & 0.076 & 0.087 & 0.072 & 0.157 & 0.029 &  \bf0.013 \\
		\hline
		7 & 0.091 & 0.052 & 0.070 & 0.080 & 0.072 & 0.149 & 0.027 &  \bf0.012 \\
		\hline
		8 & 0.093 & 0.039 & 0.071 & 0.086 & 0.043 & 0.164 & 0.020 &  \bf0.009 \\
		\hline
		9 & 0.093 & 0.038 & 0.071 & 0.086 & 0.043 & 0.156 & 0.019 &  \bf0.007 \\
		\hline
		10 & 0.092 & 0.040 & 0.067 & 0.084 & 0.041 & 0.147 & 0.019 &  \bf0.010 \\
		\hline
		11 & 0.093 & 0.041 & 0.068 & 0.087 & 0.042 & 0.158 & 0.020 &  \bf0.010 \\
		\hline
		12 & 0.096 & 0.042 & 0.073 & 0.088 & 0.043 & 0.157 & 0.020 &  \bf0.010 \\
		\hline
		13 & 0.093 & 0.044 & 0.072 & 0.085 & 0.040 & 0.162 & 0.020 &  \bf0.011 \\
		\hline
		14 & 0.097 & 0.045 & 0.075 & 0.088 & 0.044 & 0.168 & 0.022 & \bf0.010 \\
		\hline
		Ave. & 0.094 & 0.047 & 0.070 & 0.084 & 0.055 & 0.153 & 0.023 & \bf0.011 \\
		\hline
		\end{tabular}
	\end{table} 	
\begin{table}[!htp]
	\renewcommand{\arraystretch}{1.3}
	\caption{calculated p-values for the probability of error, snr, cnr and fc analysis using different tested methods.}
	\label{Ta9}
	\centering
	\tabcolsep 2.5pt
	\scriptsize
	\begin{tabular}{|c||c||c||c||c|}
		\hline
		\bf P-Value & \bf Probability of Error & \bf SNR & \bf CNR & \bf FC\\		
		\hline
		\bf SCIC & $8.81\mbox{\textsc{E}}-013$& $5.07\mbox{\textsc{E}}-04$& $2.48\mbox{\textsc{E}}-04$ &$1.6\mbox{\textsc{E}}-02$\\
		\hline
		\bf SGD & $1.3\mbox{\textsc{E}}-07$& $1.76\mbox{\textsc{E}}-04$& $6.10\mbox{\textsc{E}}-01$ &$1.45\mbox{\textsc{E}}-01$\\
		\hline
		\bf LS & $4.90\mbox{\textsc{E}}-14$& $3.1\mbox{\textsc{E}}-02$& $7.17\mbox{\textsc{E}}-01$ &$2.18\mbox{\textsc{E}}-02$\\
		\hline
		\bf LEMS & $2.95\mbox{\textsc{E}}-010$& $5.12\mbox{\textsc{E}}-03$& $1.15\mbox{\textsc{E}}-04$ & $8.56\mbox{\textsc{E}}-03$\\
		\hline
		\bf BCFCM & $4.1\mbox{\textsc{E}}-017$& $2.18\mbox{\textsc{E}}-06$& $5.04\mbox{\textsc{E}}-04$ &$4.36\mbox{\textsc{E}}-04$\\
		\hline
		\bf MCBC & $1.77\mbox{\textsc{E}}-012$& $7.3\mbox{\textsc{E}}-05$& $1.43\mbox{\textsc{E}}-02$ &$4.2\mbox{\textsc{E}}-02$\\
		\hline
		\bf NCBC &$\bf 6.15\mbox{\textsc{E}}-019$& $\bf7.85\mbox{\textsc{E}}-08$& $\bf3.02\mbox{\textsc{E}}-05$ &$\bf1.41\mbox{\textsc{E}}-04$\\
		\hline
	\end{tabular}
\end{table}

\subsubsection{Visual Assessment}
Sample slices of reconstructed DW-MR data using different tested methods are shown in Fig.~\ref{fig5} and Fig.~\ref{fig6}, with Fig.~\ref{fig5} acquired at b-value=100 $s/mm^{2}$ and Fig.~\ref{fig6} acquired at b-value=1000 $s/mm^{2}$. It can be observed that while all methods could suppress the effect of strong bias field near the ERC to a certain degree, there remains noticeable residual bias field artifacts in the results produced by all methods except the proposed NCBC method. Furthermore, the visual assessment of these results indicates that the proposed NCBC method was able to provide improved bias correction when compared to the other tested approaches for both low b-value and high b-value acquisitions, while suppressing data noise at the same time. This is most evident by the improved intensity homogeneity in the prostate region for the reconstructed DW-MR data using the proposed NCBC method. As such, using the proposed NCBC method could help to improve the overall visualization of different prostate structures compared to the uncorrected DW-MR data as the effective bias field associated with the ERC is greatly suppressed.
To see the effect of proposed NCBC method for the aim of tumor visualization, all methods are also tested on the generated ADC maps for the two patients with prostate tumor  as the reconstructed images are shown in Fig.~\ref{fig7} and Fig.~\ref{fig8}. It can be observed that using the proposed NCBC method improves the image contrast such that facilitates the distinguishing of tumor region from normal tissues. These visual results further reinforce the quantitative results in illustrating the effectiveness of the proposed NCBC method.

\subsection{Computational Performance Analysis}
All tested methods were implemented using MATLAB and C++ code and tested on an AMD Athlon II X3 3.10 GHz machine with 12GB of RAM. The computation time of each method was calculated and reported in Table \ref{Ta10}. As the timing process results show, the SCIC method had the lowest average computation time of $0.142$ seconds, with the LEMS method having the highest average computation time of $76.448$ seconds. The proposed NCBC method, where one should note that the underlying implementation has not been optimized, has an average computation time of $14.424$ seconds.

\begin{table}[!htp]
	\scriptsize
	\renewcommand{\arraystretch}{1.3}
	\caption{computation times for all tested methods.}
	\label{Ta10}
	\centering
	\tabcolsep 2.5pt
	\begin{tabular}{cccccccc}
		\hline
		\bf Case & \bf SCIC & \bf SGD & \bf LS & \bf LEMS & \bf BCFCM & \bf MCBC& \bf NCBC\\
		\hline
		1 & \bf0.123 & 1.023 & 11.640 & 75.261 & 56.171 & 4.231 & 14.651 \\
		\hline
		2 & \bf0.156 & 1.058 & 10.230 & 72.112 & 50.231 & 3.651 & 14.681 \\
		\hline
		3 & \bf0.114 & 1.121 & 11.030 & 74.112 & 53.987 & 2.311 & 15.312 \\
		\hline
		4 & \bf0.132 & 1.103 & 10.280 & 70.123 & 50.123 & 4.711 & 14.985 \\
		\hline
		5 & \bf0.113 & 1.210 & 11.220 & 80.123 & 60.101 & 4.231 & 15.231 \\
		\hline
		6 & \bf0.145 & 1.350 & 11.050 & 82.123 & 55.411 & 4.511 & 14.007 \\
		\hline
		7 & \bf0.192 & 1.010 & 10.390 & 79.235 & 53.112 & 4.326 & 13.987 \\
		\hline
		8 & \bf0.125 & 1.192 & 12.350 & 78.123 & 52.114 & 4.789 & 14.055 \\
		\hline
		9 & \bf0.112 & 1.230 & 11.087 & 52.311 & 45.321 & 3.965 & 13.256 \\
		\hline
		10 & \bf0.113 & 1.021 & 10.982 & 85.123 & 51.111 & 5.311 & 14.521 \\
		\hline
		11 & \bf0.156 & 1.052 & 11.140 & 89.156 & 50.123 & 4.120 & 14.008 \\
		\hline
		12 & \bf0.204 & 1.032 & 11.380 & 69.235 & 52.123 & 4.155 & 13.251 \\
		\hline
		13 & \bf0.182 & 1.024 & 12.010 & 93.112 & 57.564 & 4.369 & 15.011 \\
		\hline
		14 & \bf0.127 & 1.056 & 10.235 & 70.120 & 60.231 & 4.659 & 14.978 \\
		\hline
		Ave. & \bf0.142 & 1.106 & 11.073 & 76.448 & 53.409 & 4.239 & 14.424 \\
		\hline
	\end{tabular}
\end{table}
\vspace{-0.1em}
There are a number of limitations with the proposed NCBC method that should be accounted in the future direction of this work. First, the proposed framework only takes into account bias field effects; compensating for the other sources of degradation such as motion degradations as well within the framework could lead to a great quality improvement in reconstructed MR images. Second, the proposed framework does not take into account any prior information about the imaged tissue structure and characteristics, nor does it take into account prior information about the endorectal coil being used, which can be highly useful for improving image quality.  Incorporating such additional systematic and tissue information into the proposed framework is left for the future direction of this work.

\section{Conclusions}
\label{conclusion}
In this work, a novel noise-compensated, bias-corrected reconstruction method was proposed for DW-MR imaging. The proposed method takes advantage of a novel stochastically fully connected joint conditional random field model to simultaneously handle the effects of intensity inhomogeneities and data noise present in DW-MR data. The proposed NCBC technique has a potential for being integrated into different computational DW-MR imaging frameworks for the aim of providing better visualization of the whole prostate gland. The proposed technique shows improved results compared to the previously proposed bias correction algorithms in terms of both improving DW-MR data quality as well as consistency of the results. With using the proposed NCBC method, the results are improved for all synthetic phantom, physical phantom, as well as real DW-MR experiments. As a future work, it would be interesting to evaluate whether the the quality improvement of DW-MR images reconstructed using the proposed NCBC method can lead to a better separability of cancerous and healthy tissue and specifically at the earlier stages of the prostate cancer. While the preliminary results using all tested datasets is promising for providing better visualization of the prostate tissue, further investigation using different clinical datasets will be needed to confirm this aspect in detail. In addition, the sensitivity analysis of the proposed NCBC method against the amount of MR noise as well as intensity inhomogeneity needs to be investigated in the future studies of this work.

%
%
%
%
%

\vspace{-0.15in}
\section*{AUTHOR CONTRIBUTIONS}
\vspace{-0.05in}
A. B, M.J .S, and A. W contributed to the design and implementation of the concept. A. B, M.J. S, F. K, and A. W contributed to the design and implementation of the experiments, and performing statistical analysis. F. K and M. H were involved in collecting and reviewing the data. All authors contributed to the writing and reviewing of the paper.

\vspace{-0.15in}
\section*{Acknowledgment}
\vspace{-0.05in}
This research is supported by Ontario Institute of Cancer Research (OICR), Canada Research Chairs, Natural Sciences and Engineering Research Council of Canada (NSERC), and Ministry of Research and Innovation of Ontario.

\bibliographystyle{IEEEtran}
\addcontentsline{toc}{chapter}{\textbf{References}}

\bibliography{sample}


\end{document}